\documentstyle [12pt] {article}

\catcode`\@=11
\def\@citex[#1]#2{%
\if@filesw \immediate \write \@auxout {\string \citation {#2}}\fi
\@tempcntb\m@ne \let\@h@ld\relax \def\@citea{}%
\@cite{%
  \@for \@citeb:=#2\do {%
    \@ifundefined {b@\@citeb}%
      {\@h@ld\@citea\@tempcntb\m@ne{\bf ?}%
      \@warning {Citation `\@citeb ' on page \thepage \space
undefined}}%
      {\@tempcnta\@tempcntb \advance\@tempcnta\@ne%
      \@tempcntb\number\csname b@\@citeb \endcsname \relax%
      \ifnum\@tempcnta=\@tempcntb 
	\ifx\@h@ld\relax%
	  \edef \@h@ld{\@citea\csname b@\@citeb\endcsname}%
	\else%
	  \edef\@h@ld{\ifmmode{-}\else--\fi\csname
b@\@citeb\endcsname}%
	\fi%
      \else
	\@h@ld\@citea\csname b@\@citeb \endcsname%
	\let\@h@ld\relax%
      \fi}%
    \def\@citea{,\penalty\@highpenalty\,}%
  }\@h@ld
}{#1}}

\def\@citeb#1#2{{[#1]\if@tempswa , #2\fi}}
\def\@citeu#1#2{{$^{#1}$\if@tempswa , #2\fi }}
\def\@citep#1#2{{#1\if@tempswa , #2\fi}}

\def\bcites{         
	\catcode`\@=11
	\let\@cite=\@citeb
	\catcode`\@=12
}

\def\upcites{         
	\catcode`\@=11
	\let\@cite=\@citeu
	\catcode`\@=12
}

\def\plaincites{      
	\catcode`\@=11
	\let\@cite=\@citep
	\catcode`\@=12
}

\newcount\hour
\newcount\minute
\newtoks\amorpm
\hour=\time\divide\hour by 60
\minute=\time{\multiply\hour by 60 \global\advance\minute by-\hour}
\edef\standardtime{{\ifnum\hour<12 \global\amorpm={am}%
	\else\global\amorpm={pm}\advance\hour by-12 \fi
	\ifnum\hour=0 \hour=12 \fi
	\number\hour:\ifnum\minute<10
0\fi\number\minute\the\amorpm}}
\edef\militarytime{\number\hour:\ifnum\minute<10
0\fi\number\minute}

\def\draftlabel#1{{\@bsphack\if@filesw {\let\thepage\relax
   \xdef\@gtempa{\write\@auxout{\string
      \newlabel{#1}{{\@currentlabel}{\thepage}}}}}\@gtempa
   \if@nobreak \ifvmode\nobreak\fi\fi\fi\@esphack}
	\gdef\@eqnlabel{#1}}
\def\@eqnlabel{}
\def\@vacuum{}
\def\marginnote#1{}
\def\draftmarginnote#1{\marginpar{\raggedright\scriptsize\tt#1}}
\def\draft{
	\pagestyle{plain}
	\overfullrule=2pt
	\oddsidemargin -.5truein
	\def\@oddhead{\sl \phantom{\today\quad\militarytime} \hfil
	\smash{\Large\sl DRAFT} \hfil \today\quad\militarytime}
	\let\@evenhead\@oddhead
	\let\label=\draftlabel
	\let\marginnote=\draftmarginnote
	\def\ps@empty{\let\@mkboth\@gobbletwo
	\def\@oddfoot{\hfil \smash{\Large\sl DRAFT} \hfil}
	\let\@evenfoot\@oddhead}

\def\@eqnnum{(\theequation)\rlap{\kern\marginparsep\tt\@eqnlabel}%
	\global\let\@eqnlabel\@vacuum}  }

\def\blackfonts{
	\font\blackboard=msbm10 scaled\magstep1
	\font\blackboards=msbm8
	\font\blackboardss=msbm6
}

\def\nblack{            
	\def\ZZ{{Z \n{10} Z}}
	\def\NN{{N \n{14} N}}
	\def\CC{{C \n{11} C}}
	\def\RR{{R \n{11} R}}
	\def\QQ{{Q \n{12} Q}}
	\def\PP{{P \n{11} P}}
}

\def\prep{         
	\catcode`\@=11
	\input art10.sty
	\catcode`\@=12
	
	\let\small\null
	\def\blackfonts{
		\font\blackboard=msbm10
		\font\blackboards=msbm7
		\font\blackboardss=msbm5
	}
	\let\sl\it
	\twocolumn
	\sloppy
	\voffset=-2.54truecm
	\hoffset=-2.54truecm
	\flushbottom
	\parindent 1em
	\leftmargini 2em
	\leftmarginv .5em
	\leftmarginvi .5em
	\marginparwidth 48pt
	\marginparsep 10pt
	\setlength{\columnsep}{2truecm}
	\setlength{\textwidth}{25.4truecm}
	\setlength{\textheight}{17truecm}
>	\baselineskip=16pt
	\oddsidemargin .18truein
	\evensidemargin .17truein
}

\def\eqalign#1{\null\,\vcenter{\openup\jot\m@th
  \ialign{\strut\hfil$\displaystyle{##}$&$\displaystyle{{}##}$\hfil
      \crcr#1\crcr}}\,}
\def\eqalignno#1{\displ@y \tabskip\centering
  \halign
to\displaywidth{\hfil$\@lign\displaystyle{##}$\tabskip\z@skip
    &$\@lign\displaystyle{{}##}$\hfil\tabskip\centering
    &\llap{$\@lign##$}\tabskip\z@skip\crcr
    #1\crcr}}

\def\section{\@startsection {section}{1}{\z@}{3.ex plus 1ex minus
 .2ex}{2.ex plus .2ex}{\large\bf}}
\def\subsection{\@startsection{subsection}{2}{\z@}{2.75ex plus 1ex
minus
 .2ex}{1.5ex plus .2ex}{\bf}}

\def\appendix{{\newpage\section*{Appendix}}\let\appendix\section%
	{\setcounter{section}{0}
	\gdef\thesection{\Alph{section}}}\section}

\def\abstract{\if@twocolumn
\section*{Abstract}
\else 
\begin{center}
{\bf Abstract\vspace{-.5em}\vspace{0pt}}
\end{center}
\quotation
\fi}

\catcode`\@=12

\def\ibid{{\it ibid.\/}}

\newcommand{\beq}{\begin{equation}}
\newcommand{\eeq}{\end{equation}}
\newcommand{\beqa}{\begin{eqnarray}}
\newcommand{\eeqa}{\end{eqnarray}}

\newcommand{\cN}{{\cal N}}

\def\noj#1,#2,{{\bf #1} (19#2)\ }
\def\jou#1,#2,#3,{{\sl #1\/ }{\bf #2} (19#3)\ }
\def\ann#1,#2,{{\sl Ann.\ Physics\/ }{\bf #1} (19#2)\ }
\def\cmp#1,#2,{{\sl Comm.\ Math.\ Phys.\/ }{\bf #1} (19#2)\ }
\def\ma#1,#2,{{\sl Math.\ Ann.\/ }{\bf #1} (19#2)\ }
\def\jd#1,#2,{{\sl J.\ Diff.\ Geom.\/ }{\bf #1} (19#2)\ }
\def\invm#1,#2,{{\sl Invent.\ Math.\/ }{\bf #1} (19#2)\ }
\def\cq#1,#2,{{\sl Class.\ Quantum Grav.\/ }{\bf #1} (19#2)\ }
\def\cqg#1,#2,{{\sl Class.\ Quantum Grav.\/ }{\bf #1} (19#2)\ }
\def\ijmp#1,#2,{{\sl Int.\ J.\ Mod.\ Phys.\/ }{\bf A#1} (19#2)\ }
\def\jmphy#1,#2,{{\sl J.\ Geom.\ Phys.\/ }{\bf #1} (19#2)\ }
\def\jams#1,#2,{{\sl J.\ Amer.\ Math.\ Soc.\/ }{\bf #1} (19#2)\ }
\def\grg#1,#2,{{\sl Gen.\ Rel.\ Grav.\/ }{\bf #1} (19#2)\ }
\def\mpl#1,#2,{{\sl Mod.\ Phys.\ Lett.\/ }{\bf A#1} (19#2)\ }
\def\nc#1,#2,{{\sl Nuovo Cim.\/ }{\bf #1} (19#2)\ }
\def\np#1,#2,{{\sl Nucl.\ Phys.\/ }{\bf B#1} (19#2)\ }
\def\pl#1,#2,{{\sl Phys.\ Lett.\/ }{\bf #1B} (19#2)\ }
\def\pla#1,#2,{{\sl Phys.\ Lett.\/ }{\bf #1A} (19#2)\ }
\def\pr#1,#2,{{\sl Phys.\ Rev.\/ }{\bf #1} (19#2)\ }
\def\prd#1,#2,{{\sl Phys.\ Rev.\/ }{\bf D#1} (19#2)\ }
\def\prl#1,#2,{{\sl Phys.\ Rev.\ Lett.\/ }{\bf #1} (19#2)\ }
\def\prp#1,#2,{{\sl Phys.\ Rept.\/ }{\bf #1C} (19#2)\ }
\def\ptp#1,#2,{{\sl Prog.\ Theor.\ Phys.\/ }{\bf #1} (19#2)\ }
\def\ptpsup#1,#2,{{\sl Prog.\ Theor.\ Phys.\/ Suppl.\/ }{\bf #1}
(19#2)\ }
\def\rmp#1,#2,{{\sl Rev.\ Mod.\ Phys.\/ }{\bf #1} (19#2)\ }
\def\yadfiz#1,#2,#3[#4,#5]{{\sl Yad.\ Fiz.\/ }{\bf #1} (19#2) #3%
\ [{\sl Sov.\ J.\ Nucl.\ Phys.\/ }{\bf #4} (19#2) #5]}
\def\zh#1,#2,#3[#4,#5]{{\sl Zh..\ Exp.\ Theor.\ Fiz.\/ }{\bf #1}
(19#2) #3%
\ [{\sl Sov.\ Phys.\ JETP\/ }{\bf #4} (19#2) #5]}

\def\beq{\begin{equation}}
\def\eeq{\end{equation}}
\def\beqar{\begin{eqnarray}}
\def\eeqar{\end{eqnarray}}
\def\non{\nonumber}

\def\nfrac#1#2{{\displaystyle{\vphantom1\smash{\lower.5ex\hbox{\sma
ll$#1$}}%
	\over\vphantom1\smash{\raise.25ex\hbox{\small$#2$}}}}}

\def\p#1{\mskip#1mu}
\def\n#1{\mskip-#1mu}
\def\stop{\p6.}
\def\comma{\p6,}

\def\lae{\mathrel{\mathop{\smash{\lower .5 ex
\hbox{$\stackrel<\sim$}}}}}
\def\lae{\mathrel{\mathop{\smash{\lower .5 ex
\hbox{$\stackrel>\sim$}}}}}

\def\Tr{{\rm Tr}}
\def\l:{\mathopen{:}\,}
\def\r:{\,\mathclose{:}}

\catcode`\@=11
\def\theequation{\arabic{equation}}
\catcode`\@=12

\nblack
\bcites

\nblack

\catcode`\@=11
\def\theequation{\thesection.\arabic{equation}}
\@addtoreset{equation}{section}
\@addtoreset{footnote}{section}
\@addtoreset{footnote}{subsection}
\catcode`\@=12

\newcommand{\drawsquare}[2]{\hbox{%
\rule{#2pt}{#1pt}\hskip-#2pt
\rule{#1pt}{#2pt}\hskip-#1pt
\rule[#1pt]{#1pt}{#2pt}}\rule[#1pt]{#2pt}{#2pt}\hskip-#2pt
\rule{#2pt}{#1pt}}

\newcommand{\Yfund}{\raisebox{-.5pt}{\drawsquare{6.5}{0.4}}}

\def\vbr{\vphantom{\sqrt{F_e^i}}}

\def\dim{{\rm dim}}

\newcommand{\beqn}{\begin{equation}}
\newcommand{\eeqn}{\end{equation}}
\newcommand{\beqnarray}{\begin{eqnarray}}
\newcommand{\eeqnarray}{\end{eqnarray}}
\newcommand {\bear} [1] {\begin {array} {#1}}
\newcommand {\ear} {\end {array}}

\newcommand {\beqarn} {\begin{eqnarray*}}
\newcommand {\eeqarn} {\end{eqnarray*}}

\newcommand {\mrm} [1] {\mathrm {#1}}

\newcommand {\myref} [1]	%
	{%
	\begin{thebibliography} {99}	%
			{#1}	%
	\end {thebibliography}}

\def\emph#1{{\em #1}}
\def\mathbf#1{{\bf #1}}
\def\mathrm#1{{\rm #1}}
\def\mathit#1{{\it #1}}

\catcode`\@=11
\@addtoreset{footnote}{section}
\catcode`\@=12

\newcounter	{exinsert}	[subsection]

\renewcommand {\theexinsert}	%
{	\thesubsection.\arabic{equation}}

\renewcommand {\theequation} {\thesection.\arabic{equation}}

\catcode`\@=11
\@addtoreset{equation}{section}
\catcode`\@=12

\newenvironment {exinsert} [1]	%
{	
	\begin {quotation}	%
	\refstepcounter {equation}	%
	{\bf {#1}} \theequation. \,\,	%
}
{	
	\end {quotation}
}

\newcommand {\bprop}
{\begin {exinsert} {Proposition}
}
\newcommand {\eprop} {\end {exinsert} }

\newcommand {\bexe} {\begin {exinsert} {Exercise}}
\newcommand {\eexe} {\end {exinsert} }

\newcommand {\bexa} {\begin {exinsert} {Example}}
\newcommand {\eexa} {\end {exinsert} }

\newcommand {\literal} [1] {{ \mbox {$\mrm {#1}$}}}

\newcommand {\grad} {\nabla}

\newcommand {\ncmd} [2] {\newcommand {#1} {#2}}

\ncmd {\vgrad} {{\vec \grad}}

\ncmd {\Mv} {{\cal M}_V}
\ncmd {\Mh} {{\cal M}_H}

\ncmd {\adj} {\literal {adj}}	
\ncmd {\Mp} {M_\literal {planck}}	

\ncmd {\x} {$\times$}

\newcommand {\btab} [3]
  {\begin{table} [htb]
   \begin{center}
   \caption {#2}	\label {tab:#1}
   \begin {tabular} {#3}
   \hline}

\newcommand {\etab}
	{  \hline   \end {tabular}
   \end{center}
  \end{table}}

\ncmd {\bR} {\textbf {R}}
\ncmd {\bC} {\textbf {C}}

\ncmd {\CQ} {\cal Q}

\def\jou#1,#2,#3,{{\sl #1\/ }{\bf #2} (19#3)\ }
\def\ibid#1,#2,{{\sl ibid.\/ }{\bf #1} (19#2)\ }
\def\am#1,#2,{{\sl Acta. Math.\/ } {\bf #1} (19#2)\ }
\def\annp#1,#2,{{\sl Ann.\ Physics\/ }{\bf #1} (19#2)\ }
\def\cmp#1,#2,{{\sl Comm.\ Math.\ Phys.\/ }{\bf #1} (19#2)\ }
\def\cqg#1,#2,{{\sl Class.\ Quantum Grav.\/ }{\bf #1} (19#2)\ }
\def\dm#1,#2,{{\sl Duke\ Math.\ J.\/ }{\bf #1} (19#2)\ }
\def\ijmpa#1,#2,{{\sl Int.\ J.\ Mod.\ Phys.\/ }{\bf A#1} (19#2)\ }
\def\invm#1,#2,{{\sl Invent.\ Math.\/ }{\bf #1} (19#2)\ }
\def\jhep#1,#2,{{\sl JHEP\/ }{\bf #1} (19#2)\ }
\def\jmp#1,#2,{{\sl J.\ Geom.\ Phys.\/ }{\bf #1} (19#2)\ }
\def\jams#1,#2,{{\sl J.\ Diff.\ Geom.\/ }{\bf #1} (19#2)\ }
\def\jdg#1,#2,{{\sl J.\ Amer.\ Math.\ Soc.\/ }{\bf #1} (19#2)\ }
\def\jpa#1,#2,{{\sl J.\ Phys.\ A.\/ }{\bf #1} (19#2)\ }
\def\grg#1,#2,{{\sl Gen.\ Rel.\ Grav.\/ }{\bf #1} (19#2)\ }
\def\mpla#1,#2,{{\sl Mod.\ Phys.\ Lett.\/ }{\bf A#1} (19#2)\ }
\def\ma#1,#2,{{\sl Math.\ Ann.\/ } {\bf #1} (19#2)\ }
\def\nc#1,#2,{{\sl Nuovo\ Cim.\/ }{\bf #1} (19#2)\ }
\def\npb#1,#2,{{\sl Nucl.\ Phys.\/ }{\bf B#1} (19#2)\ }
\def\plb#1,#2,{{\sl Phys.\ Lett.\/ }{\bf #1B} (19#2)\ }
\def\pla#1,#2,{{\sl Phys.\ Lett.\/ }{\bf #1A} (19#2)\ }
\def\pnas#1,#2,{{\sl Proc.Nat.Acad.Sci.\/ }{\bf #1} (19#2)\ }
\def\prev#1,#2,{{\sl Phys.\ Rev.\/ }{\bf #1} (19#2)\ }
\def\prd#1,#2,{{\sl Phys.\ Rev.\/ }{\bf D#1} (19#2)\ }
\def\prl#1,#2,{{\sl Phys.\ Rev.\ Lett.\/ }{\bf #1} (19#2)\ }
\def\prpt#1,#2,{{\sl Phys.\ Rept.\/ }{\bf #1C} (19#2)\ }
\def\ptp#1,#2,{{\sl Prog.\ Theor.\ Phys.\/ }{\bf #1} (19#2)\ }
\def\ptpsup#1,#2,{{\sl Prog.\ Theor.\ Phys.\/ Suppl.\/ }{\bf #1} (19#2)\ }
\def\rmp#1,#2,{{\sl Rev.\ Mod.\ Phys.\/ }{\bf #1} (19#2)\ }
\def\sm#1,#2,{{\sl Selec. Math.\/ }{\bf #1} (19#2)\ }
\def\yadfiz#1,#2,#3[#4,#5]{{\sl Yad.\ Fiz.\/ }{\bf #1} (19#2) #3%
\ [{\sl Sov.\ J.\ Nucl.\ Phys.\/ }{\bf #4} (19#2) #5]}
\def\zh#1,#2,#3[#4,#5]{{\sl Zh.\ Exp.\ Theor.\ Fiz.\/ }{\bf #1} (19#2) #3%
\ [{\sl Sov.\ Phys.\ JETP\/ }{\bf #4} (19#2) #5]}

\begin {document}


\begin{titlepage}

\begin{center}
\today
\hfill LBNL-41562, UCB-PTH-98/16\\
\hfill                  hep-th/9803167

\vskip 1.5 cm
{\large \bf Orbifolds of $AdS_5\times S^{5}$
and $4d$ Conformal Field Theories}
\vskip 1 cm
{Yaron Oz \footnote {email: yaronoz@thsrv.lbl.gov} and
John Terning \footnote {email: terning@alvin.lbl.gov}}\\
\vskip 0.5cm
{\sl Department of Physics,
University of California at Berkeley\\
366 Le\thinspace Conte Hall, Berkeley, CA 94720-7300, U.S.A.\\
and\\
Theoretical Physics Group, Mail Stop 50A--5101\\
Ernest Orlando Lawrence Berkeley National Laboratory, \\
Berkeley, CA 94720, U.S.A.\\}

\end{center}

\vskip 0.5 cm
\begin{abstract}
We study the relation between the large $N$ limit of four  dimensional
${\cal N}=2,1,0$ conformal field theories and supergravity on orbifolds of
$AdS_5\times S^{5}$.
We analyze the the Kaluza-Klein
states of the supergravity theory and relate them to the
spectrum of (chiral) primary operators of
the (super) conformal field theories.

\end{abstract}
\end{titlepage}

\section{Introduction}

Recently a duality  between  superconformal field
theories (SCFTs) in $d$ dimensions and string or M theory compactified
on anti-de Sitter (AdS)
spaces of the form $AdS_{d+1} \times W$ has  been proposed
in \cite{mal} (see also
\cite{DPS,kleb1,kleb2,kleb3,kleb4,MS,Polyakov,FF}).  Here $W$ is a
compact manifold which in the maximally supersymmetric cases is a
sphere.  A precise correspondence between the supergravity limit on
the $AdS_{d+1}$ side and an appropriate large $N$ limit on the
conformal field theory side has been formulated in
\cite{GKP,witten-one}\footnote{See also \cite{FF} for the relation between bulk fields and
boundary composites.}.  According to \cite{witten-one} the correlation
functions in the conformal field theory, which has as its spacetime
$M_d$, the boundary at infinity of $AdS_{d+1}$, can be calculated
via the dependence of the supergravity action on the
asymptotic behavior of its fields at the boundary $M_d$. In
particular, one can deduce the scaling dimensions of operators in the
conformal field theory from the masses of particles in string theory
(or M theory). Using this correspondence, the dimensions of chiral
primary operators
in four dimensional ${\cal N}=4$ super-Yang-Mills (SYM)
were matched with the
masses of Kaluza-Klein states on $AdS_5 \times S^5$.
Related works which appeared recently are [12-43].

In \cite{KS,LNV} a relation between several classes of
four dimensional ${\cal N}=2,1,0$  conformal field theories
and Type IIB supergravity (string) theory on orbifolds of  $AdS_5\times S^{5}$
was proposed.
The orbifolds preserve the  $AdS_5$ structure and its isometry group $SO(4,2)$
which becomes the conformal
symmetry of the four dimensional theory. The orbifold action on $S^5$ breaks
some or all the ${\cal N}=4$
supersymmetry.
When the orbifold group $\Gamma$ acts freely on $S^5$ there is a limit where
supergravity provides
an  applicable description.
When the orbifold action is not free only the string theory description is
reliable.
Some of these  $\cN=1,2$ models were shown to be conformal \cite{KS}.
The analysis of the one-loop and two-loop $\beta$ functions of
the general orbifold  theories in  \cite{LNV} showed that they  indeed vanish.
Furthermore the analysis in \cite{BKV} shows that these theories have indeed
a fixed line (fixed hypersurface) at least in the large $N$ limit.
In the nonsupersymmetric models, the $\beta$ functions need not vanish at
finite $N$ and they do not, as we will comment
on in the discussion section.

In this paper we will study the proposed duality by analyzing the Kaluza-Klein
states of the
supergravity theory on the  orbifolds of  $AdS_5\times S^{5}$ and relating them
to the (chiral) primary
operators of the (super) conformal field theories on the boundary.
In the next section we will briefly review the relation between $AdS$
supergravity (string) theory
and SCFTs on the boundary of the $AdS$ space. In particular we will review in
detail the relation
between the Kaluza-Klein harmonics of supergravity on  $AdS_5\times S^{5}$
and the chiral primary operators of $\cN=4$ SYM in four dimensions.
In section 3 we will analyze the   Kaluza-Klein harmonics of supergravity on
orbifolds
 $AdS_5\times S^{5}/ \Gamma$ where $\Gamma \subset SU(3)$ is a discrete
subgroup.
We will relate the  Kaluza-Klein modes to the chiral primary operators of the
$\cN=1$ theory on
the boundary of the $AdS$ space.
In section 4 we will perform similar analysis when the orbifold group
 $\Gamma \subset SU(2)$. In this case one gets an $\cN=2$ theory on the
boundary
of the $AdS$ space.
However now the orbifold action is not free and one in general does not expect
supergravity
to provide an applicable description. Nevertheless we will still be able
to relate the  Kaluza-Klein modes to the chiral primary operators of the
boundary
$\cN=2$ theory.
This means
that chiral information is still reliably encoded in the supergravity
description.
In section 5 we study  a possible relation between Kaluza-Klein states and
primary operators of the boundary CFT
when the orbifold group  $\Gamma \subset
SU(4)$
and the theory on the boundary is not supersymmetric.
Section 6  is devoted to a summary of the results and discussion.

\section{SCFT/$AdS$ Relation: ${\cal N}=4$ SYM In Four Dimensions}

In this section we will briefly review the SCFT/$AdS$ relation
proposed in \cite{GKP,witten-one}.
One of the examples of this relation is between
$\cN=4$ SYM in four dimensions and Type IIB supergravity (or string) theory on
$AdS_5 \times S^5$.
This example will be of particular importance for us since
orbifolds of this relation will be studied in next sections.

The boundary $M_d$ of $AdS_{d+1}$ is a
$d$-dimensional Minkowski space with points at infinity added.  The
isometry group of $AdS_{d+1}$ is $SO(d,2)$. It is also the conformal
group on $M_d$.  The proposed duality relates
string theory (or M theory) on $AdS_{d+1}$ to the large $N$ limit of
 SCFTs on its boundary
$M_d$.  In the Euclidean version the boundary is $S^d$.
Consider the maximally supersymmetric case, so that the
internal space is also a sphere.  Let $\phi$
be a scalar field on $AdS_{d+1}$ and $\phi_0$ its restriction to the
boundary $S^d$.  According to the SCFT/$AdS$ relation $\phi_0$
couples to a conformal operator ${\cal O}$ on the
boundary via $\int_{S^d} \phi_0 {\cal O}$.

When $\phi$ has mass $m$ the corresponding operator
${\cal O}$ has conformal dimension $\Delta$ given by
\beq
m^2 = \Delta(\Delta-d) \stop
\label{dim}
\eeq
Irrelevant, marginal, and relevant operators of the boundary theory
correspond to massive, massless, and ``tachyonic'' modes in the supergravity
theory.  If a $p$-form $C$ on $AdS$ is coupled to a $d-p$ form
operator $\cal O$ on the boundary, then the relation between the mass of
$C$ and the conformal dimension of $\cal O$ is given by
\beq
m^2 = (\Delta+p)(\Delta+p-d) \stop
\label{dimp}
\eeq
The value of $m^2$ in this formula refers to the eigenvalue of the Laplace
operator on the $AdS$ space. In the supergravity literature, the values that
are usually quoted for $p$-forms are the eigenvalues ${\tilde m}^2$ of the
appropriate Maxwell-like operators. The relation of these to the dimension
is given by 
\beq
{\tilde m}^2 = (\Delta-p)(\Delta+p-d) \stop
\label{dimpm}
\eeq
Formula (\ref{dimpm}) can be derived
by repeating the analysis of \cite{witten-one} 
using the Maxwell type equations for the p-forms. Alternatively, 
one can  use the definition of the mass via the quadratic
Casimir of the $SO(d,2)$ isometry group of $AdS_{d+1}$ as was done for $AdS_5$
in \cite{FFZ}.

The massless graviton in the $AdS$ supergravity  couples
to the dimension $d$ stress-energy tensor of the SCFT.
When the internal space $W$ has continuous
rotational symmetry, there are also $AdS$ massless vector
fields in its adjoint representation which  couple to the dimension $d-1$
$R$-symmetry currents
of the SCFT.

Consider the Type IIB superstring theory on $AdS_5\times S^5$ with a 5-form
flux of
$N$ units
on $S^5$ and radius of curvature $(g_{st}N)^{\frac{1}{4}}$. In the large $N$
limit with $g_{YM}^2 N = g_{st}N$ fixed and large, string theory is
weakly coupled and the supergravity description
is applicable. The bosonic symmetry of this compactification of ten dimensional
Type IIB supergravity is $SO(4,2) \times SO(6)$.
In \cite{mal} it was proposed that $\cN=4$ SYM in four dimensions is dual to
string theory on the
above background.
The $SO(4,2)$ part of the symmetry of the supergravity theory corresponds to
the conformal symmetry of the $\cN=4$ superconformal theory.
The $SO(6)\simeq SU(4)$ part of the symmetry, which is the isometry of $S^5$,
is the
the $R$-symmetry of the superconformal theory.

$AdS$ supergravity has multiplet shortening due to the internal symmetry
generators
of its superalgebra \cite{FN,H}.
In the maximally supersymmetric case
the Kaluza-Klein excitations of supergravity fall into short representations
of supersymmetry with spins $\leq 2$, and their mass formula is protected from
quantum and string
corrections.
They couple to chiral primary operators  of four dimensional $\cN=4$ SYM on the
boundary.
Chiral operators  are in short representations of the
superconformal algebra and their dimensions are  determined in terms of the
$R$-symmetry representation  and  cannot receive any corrections
\cite{seiberg,minwalla}.
An $\cN=1$ superconformal subalgebra of the $\cN =4$  superconformal algebra
has a generator, $R$,
of the $U(1)_R$ symmetry. The dimensions of chiral  operators are
determined by their R charges
\beq
\dim({\cal O}) = {{3}\over{2}} R .
\label{R}
\eeq
Since a bulk field $\phi$ with boundary value  $\phi_0$
couples to a conformal field ${\cal O}$ on the boundary via $\int_{S^d}
\phi_0 {\cal O}$, $\phi_0$ carries opposite $R$-charge to that of  ${\cal O}$.
Multiplet shortening in $AdS$ supergravity is expected also with a
reduced number
of supersymmetries \cite{FN,H}
and we expect the mass formulas of the  Kaluza-Klein excitations
to be  protected from
quantum and string
corrections.

The spectrum of Kaluza-Klein harmonics of supergravity on  $AdS_5\times S^5$
was derived in
\cite{van0,gm}.
The Kaluza-Klein harmonics fall into
irreducible representations of $SU(4)$.
We will now review the
families that contain fields with negative or zero mass.

There is one family of spin-2 fields. The mass formula, in $1/\sqrt{\alpha'}$
units,
is given by
\beq
m^2 = k(k+4),~~~~~k \geq 0 \stop
\label{grav}
\eeq
The $SU(4)$ Dynkin labels of the representations are $(0,k,0)$,  and the
corresponding
$SU(4)$ irreducible representations are ${\bf 1,6,20^\prime,...}$.
The
$k=0$ particle is the  graviton that couples to the dimension 4  stress-energy
tensor operator
of  the $\cN=4$
SCFT theory.

There is one family of vector fields that contains massless states with  mass
formula
 given by
\beq
m^2 = (k-1)(k+1),~~~~~k \geq 1 \stop
\label{vec}
\eeq
The Dynkin labels are $(1,k-1,1)$ and the
irreducible representations are ${\bf 15},{\bf 64},{\bf 175},...$.
The massless vector bosons at $k=1$ transform in the adjoint of
$SU(4)$ and couple to the
$SU(4)$ $R$-symmetry currents of  the $\cN=4$
SCFT theory.

There are three families of scalar fields that contain negative and massless
states.
The first family has mass formula
\beq
m^2 = k(k-4),~~~~~k \geq 2 \comma
\label{k1}
\eeq
with Dynkin labels  $(0,k,0)$
corresponding to the
irreducible representations  ${\bf 20'},{\bf 50},{\bf 105},...$.
They couple to dimension  $\Delta = k$ chiral primary operators of  the $\cN=4$
SCFT theory\footnote{More precisely, since
the  ${\bf 20^\prime}$ representation of $SU(4)$ decomposes into
representations of $SU(3) \times U(1)_R$ as
${\bf 20^\prime} =  {\bf 6}_{4/3} + \overline{\bf 6}_{-4/3} + {\bf 8}_0$
then the  ${\bf 6}_{4/3}$ Kaluza-Klein states couple to chiral primary
operators, the
 ${\bf 6}_{-4/3}$  couple to anti-chiral primary operator and the
${\bf 8}_0$ couple to operators which are neither chiral nor anti-chiral.
Nevertheless, since they all sit in the same $\cN=4$ multiplet they are all
protected from quantum corrections.}
which were identified in \cite{witten-one} as the symmetrized traceless
$\Tr(\Phi^{1_1}...\Phi^{i_k})$ of the adjoint chiral superfields
$\Phi^i,i=1,2,3$.
These operators indeed transform in the same
symmetric traceless representations of
$SU(4)$ as the Kaluza-Klein
particles (\ref{k1}).

The second family has mass formula
\beq
m^2 = (k-1)(k+3),~~~~~k \geq 0 \comma
\label{k2}
\eeq
with Dynkin labels $(0,k,2)$
corresponding to the
irreducible representations  ${\bf 10},{\bf 45},{\bf 126},...$.
They couple to dimension  $\Delta = k+3$ chiral primary operators of  the
$\cN=4$
SCFT theory
$Tr(W_{\alpha}W^{\alpha}\Phi^{1_1}...\Phi^{i_k})$ where $W_{\alpha}$ is the
field strength chiral
superfield \cite{witten-one}

The third family has mass formula
\beq
m^2 = k(k+4),~~~~~k \geq 0 \comma
\label{k3}
\eeq
with Dynkin labels  $(0,k,0)$
corresponding to the
irreducible representations  ${\bf 1},{\bf 6},{\bf 20'},...$.
The massless particle in this family ($k=0$) is the dilaton.
It couples to $Tr F^2$ in the $\cN=4$ theory.
The particles couple to  dimension $\Delta = k+4$
chiral primary operators of the boundary theory $Tr a^k F^2+...$ where $a$ is
the complex scalar in
the $\cN=2$ vector multiplet (when viewing $\cN=4$ as $\cN=2$ with
a hypermultiplet in the adjoint)
\cite{witten-one}.

The different towers of Kaluza-Klein harmonics are related by the action of the
supersymmetry generators and
can be organized in an $\cN=4$ supertower \cite{FFZ}.
For instance, the graviton, the   ${\bf 15}$ massless vector bosons and the
scalars in the above three
families in the representations
${\bf 20'},{\bf 10},{\bf 1}$ of $SU(4)$ are in the same multiplet.

\section{${\cal N}=1$ Supersymmetric Theories}

In \cite{KS,LNV} ${\cal N}=1$ SCFTs were constructed by studying D3 branes at
orbifold singularities
of the form $R^6/\Gamma$, where $\Gamma \subset SU(3)$ is a discrete subgroup.
The worldvolume theory is constructed by taking $N|\Gamma|$ D3 branes on the
covering space
and performing a projection $\Gamma$ on the worldvolume fields and the Chan
Paton
factors \cite{dm,jom,dgm}.
Conformal field theories are expected when  the representation of $\Gamma$
acting
on the Chan Paton factors
is chosen to be the $N$-fold copy of the regular representation.
In the framework  of \cite{mal} this translates to the study of Type IIB string
theory on
an orbifold of $AdS_5 \times S^5$ where the orbifold group acts only on the
$S^5$ factor.
The $SO(4,2)$ isometry of $AdS_5$  is not broken and  corresponds to the
conformal symmetry of the SCFT on the D3 branes worldvolume.
The $SO(6) \simeq SU(4)$ isometry of $S^5$  is broken to $U(1)_R \times
\Gamma$.
The $U(1)_R$ factor is the $R$-symmetry of the boundary ${\cal N}=1$ D3
brane theory.
The $\Gamma$ factor 
becomes a discrete global symmetry of the D3 brane theory\footnote{$\Gamma$ is a symmetry of the quiver
diagram description.}.  

Consider first the $Z_3$ orbifold example of \cite{KS}
\beq
X^{1,2,3} \rightarrow e^{\frac{2 \pi i}{3}} X^{1,2,3}
\label{z3}
\comma
\eeq
where $X^i$ parametrize the $R^6$ transverse to the D3 branes worldvolume.
The gauge group, global symmetries and field content of the D3 brane
theory is given in Table 1.
\begin{table}[htbp]
\centering
\begin{tabular}{cccc|c}
 & $SU(N)$ & $SU(N)$  & $SU(N)$ & $U(1)_R$

\\
\hline
$U^i$ & \Yfund & $\overline{\Yfund}$ & {\bf 1}  &  ${{2}\over{3}} \vbr$ \\
$V^i$ & {\bf 1} & \Yfund &  $\overline{\Yfund}$ &  ${{2}\over{3}} \vbr$ \\
$W^i$ & $\overline{\Yfund}$ & {\bf 1} &\Yfund &  ${{2}\over{3}} \vbr$ \\
\end{tabular}
\label{N=1}
\parbox{4in}{\caption{Field content of the ${\cal N}=1$ theory, where
$i=1,2,3$. The $SU(3)$ global symmetry is broken by the superpotential.}}
\end{table}

The orbifold (\ref{z3}) has a fixed point at the origin. Since the volume of
$S^5$ is not zero,
the orbifold action is free and the resulting manifold is smooth.
In the large $N$ limit as specified in the $\cN=4$ SYM theory in the previous
section the supergravity description is applicable.
If the relation between supergravity and SCFT of \cite{mal,witten-one} holds
also here
we expect to find  Kaluza-Klein harmonics of supergravity on $AdS_5 \times
S^5/Z_3$ corresponding to the
chiral primary operators in the $\cN=1$ theory.
In the following we will study this correspondence.
We will analyze the Kaluza-Klein harmonics of the supergravity theory and the
relation to chiral primary
operators of the boundary SCFT.

The  Kaluza-Klein harmonics of supergravity on $AdS_5 \times S^5/Z_3$  are
$Z_3$ invariant
states and can be
obtained by a $Z_3$ projection of  the
Kaluza-Klein harmonics  on $AdS_5 \times S^5$ discussed in the previous
section.
Consider  the scalar modes with masses given by (\ref{k1}).
Let us  explicitly check the relevant and marginal
chiral primary operators in this family. The  $k=2$ Kaluza-Klein particle in
(\ref{k1})
transforms in  the ${\bf 20^\prime}$ of $SU(4)$.
Decomposing \cite{slansky}
the ${\bf 20^\prime}$ into representations of $SU(3) \times U(1)_R$ gives:
\beq
{\bf 20^\prime} =  {\bf 6}_{4/3} + \overline{\bf 6}_{-4/3} + {\bf 8}_0 \stop
\label{20}
\eeq

We now have to perform the $Z_3$ projection on (\ref{20}).
The ${\bf 8}_0$ is invariant  under the $Z_3$
projection\footnote{The $Z_3$ acts on the ${\bf 3}$ of $SU(3)$ as
$(x^1,x^2,x^3) \rightarrow (e^{\frac{2 \pi i}{3}} x^1, e^{\frac{2 \pi i}{3}}
x^2, e^{\frac{-4 \pi i}{3}}x^3)$.
The ${\bf 8}$ is made of ${\bf 3}$ and $\overline{\bf 3}$.}. However these
Kaluza-Klein modes are  expected to couple  to
dimension 2 operators.
A dimension 2 chiral primary operator has $R$-charge\footnote{The sign of the
$R$-charge
assignments in the
decomposition is merely a convention.}   ${{4}\over{3}}$ (\ref{R}). Therefore
${\bf 8}_0$ has
the wrong $R$-charge to couple to a dimension 2 chiral operator and we do not
expect
any dimension 2 chiral primary operators
in the boundary $N=1$ SCFT.
We expect dimension 2 operators which are not chiral primary to couple
to the the Kaluza-Klein modes in the
${\bf 8}_0$.
In the $\cN=4$ case, the ${\bf 6}_{4/3},
\overline{\bf 6}_{-4/3}$, and ${\bf 8}_0$ in (\ref{20})
sit in the same supermultiplet and therefore the masses of the
${\bf 8}_0$ Kaluza-Klein states were protected; here there is no such
guarantee.


The  $k=3$ Kaluza-Klein particle in (\ref{k1})
which transforms in  the ${\bf 50}$ of $SU(4)$
 should  couple to a dimension 3  chiral primary operator.
Decomposing the ${\bf 50}$ into representations of $SU(3) \times U(1)_R$ gives:
\beq
{\bf 50} = {\bf 10}_2 + \overline{\bf 10}_{-2} + {\bf 15}_{2/3} + \overline{\bf
15}_{-2/3} \stop
\label{yuk1}
\eeq
The ${\bf 10}$  is invariant under the $Z_3$
projection, and this is the only
part in the decomposition (\ref{yuk1}) with the correct
$R$-charge to couple to a dimension 3
chiral primary operator.
We therefore expect 10 dimension 3 chiral primary operators in the boundary
SCFT
and we identify them with the ten  independent operators
$\Tr U^{i_1}V^{i_2}W^{i_3}$  symmetric in the $i_k$ indices. The antisymmetric
parts
have to be removed in order to form primary operators since they appear in
the superpotential and its derivatives.

The $k=4$ massless Kaluza-Klein particle  in (\ref{k1}) should couple to a
marginal
operator. It transforms in the ${\bf 105}$ which decomposes as:
\beq
{\bf 105} = {\bf 15'}_{8/3} + \overline{\bf 15'}_{-8/3} + {\bf 24}_{-4/3} +
\overline{\bf
24}_{4/3}  + {\bf 27}_0 \stop
\eeq
We see that the ${\bf 15'}$ has the right $R$-charge to couple to a dimension 4
chiral primary operator,
but it is not invariant under $Z_3$.  The ${\bf 27}$ is
invariant
under $Z_3$
but it's $R$-charge is not consistent with coupling to a dimension 4
chiral operator.
So there are no Kaluza-Klein harmonics in this family that can couple to
dimension
4 chiral primary operators and no such operators are expected in
the boundary SCFT.

In general we expect the $Z_3$ invariance and the $R$-charge
condition to  restrict the value of
$k$ to be a multiple of 3.
The Kaluza-Klein modes that survive in this family are
\beq
m^2 = 3n(3n-4),~~~~~n=1,2, \ldots
\comma
\eeq
and they couple to chiral primary operators
of dimensions
\beq
\dim({\cal O}) = \{3n,~~n=1,2, \ldots\} \comma
\label{dims1}
\eeq
in the boundary SCFT.
We can identify these  operators as symmetric operator
$O_n = \Tr(UVW)^n$.

Consider  next the scalar modes with masses given by (\ref{k2}).
Again we will  explicitly check the relevant and marginal
chiral primary operators in this family.
The Kaluza-Klein mode with $k=0$
transforms in the ${\bf 10}$ of $SU(4)$.
We decompose the
${\bf 10}$ into representations of $SU(3) \times U(1)_R$ as:
\beq
{\bf 10} = {\bf 1}_2   + {\bf 3}_{2/3} +  {\bf 6}_{-2/3} \stop
\eeq
The ${\bf 1}_2$ is the only component that is invariant under the $Z_3$
projection, and it is also the only component with the correct $R$-charge to
couple to a dimension 3 chiral operator.
In fact the ${\bf 1}$ component will be invariant under
any projection
that preserves ${\cal N}=1$ supersymmetry.
This Kaluza-Klein mode couples to
 the relevant operator $\sum_{i=1}^3\Tr W^i_{\alpha}W_i^{\alpha}$ where the
index $i$ enumerates  the
three gauge groups. This combination is dictated by the $Z_3$ global symmetry.
This operator
is a
linear combination of the gaugino bilinears.
As expected,  the gaugino bilinears will be
dimension
3 operators in any such theory derived from the ${\cal N}=4$ by projection.

The $k=1$ states in (\ref{k2}) transform in the ${\bf 45}$, which
decomposes\footnote{There is a typo in ref. \cite{slansky} in the decomposition
of the ${\bf 45}$ in Table 27, the ${\bf 6}$ is repeated.} as:
\beq
{\bf 45} = {\bf 3}_{8/3} + \overline{\bf 3}_{4/3} + {\bf 6}_{4/3} +
{\bf 8}_{0}  + {\bf 10}_{0}  + {\bf 15}_{-4/3} \stop
\eeq
The ${\bf 3}$ has the correct $R$-charge to couple to a dimension 4
chiral primary operator, but it is not
invariant under $Z_3$, while the ${\bf 10}$ is invariant under
$Z_3$ but it has
the wrong $R$-charge, so we do not expect  dimension 4 chiral primary operators
in
the boundary SCFT coupled to Kaluza-Klein modes of this
family.

In general we expect Kaluza-Klein modes with
\beq
m^2 = (3n-1)(3n+3),~~~~~n=0,1,2, \ldots
\eeq
coupled to chiral primary
operators in the SCFT with dimensions
\beq
\dim({\cal O}) = \{3n+3,~~n=0,1,2, \ldots\} \stop
\label{dims2}
\eeq
These operators can be identified as $O_n = \Tr W_{\alpha}W^{\alpha}(UVW)^n$
where we suppressed the
sum on the different gauge groups and the indices of the matter multiplets.
For $n$ bigger than zero these operators transform in the representation
constructed from the
product of  $\bf{10}$ with the $3n$'th rank symmetric tensors.

Consider now the third family (\ref{k3}).
The $k=0$ state, the dilaton, transforms in the  ${\bf 1}$ which is invariant
under the $Z_3$ projection.
It couples to the marginal operator $\sum_{i=1}^3Tr F_i^2$.
Evidently this result is independent of the choice of
$\Gamma$, and
the operator $\Tr F^2$
will be marginal in any theory obtained by $\Gamma$
projection on the $\cN=4$ theory.
 We only find one
marginal chiral primary operator in this family, since all higher values of $k$
couple to irrelevant
operators.
As before we expect Kaluza-Klein harmonics with
\beq
m^2 = 3n(3n+4),~~~~~n=0,1,2, \ldots
\eeq
coupled to chiral primary
operators in the SCFT with dimensions
\beq
\dim({\cal O}) = \{3n+4,~~n=0,1,2, \ldots\} \stop
\label{dims3}
\eeq

The graviton with $k=0$ in the spin 2 family (\ref{grav})
is in the   ${\bf 1}$ of $SU(4)$ and is invariant under the $Z_3$ projection
and in general under any $\Gamma \subset SU(4)$ projection.
As mentioned previously it couples to the stress-energy tensor.

The massless vector  $k=1$ in the spin-1 family
(\ref{vec}) is in the ${\bf 15}$ of $SU(4)$.  Decomposing the ${\bf 15}$
we find
\beq
{\bf 15} = {\bf 1}_0   + {\bf 3}_{-4/3} + \overline{\bf 3}_{4/3} +{\bf 8}_0
\stop
\eeq
The ${\bf 1}$ is invariant under $Z_3$, and it
has the correct charge to couple to the unbroken  $U(1)_R$ current.
The ${\bf 1}$ component will be invariant under any projection
that preserves ${\cal N}=1$ supersymmetry.  The
${\bf 8}_0$ is also invariant under $Z_3$ and has the correct $R$-charge,
these
are the $Z_3$ remnants of the broken $SU(3)$. The currents to which they couple
are not conserved, so
there is no guarantee that the masses of these Kaluza-Klein states and the
dimensions of the currents are
protected from quantum corrections.

It is straightforward to extend the analysis  to other projections
that preserve ${\cal N}=1$ supersymmetry. The discrete,
non-Abelian subgroups of
$SU(3)$ have been classified  in  \cite{discrete}.  Consider first
the group $\Delta(3)$,
the group of cyclic permutations on three objects.   Since $\Delta(3)$ is
a subgroup of  the other non-Abelian subgroups of $SU(3)$,
it is easy to see that using these more
complicated projections can only further restrict the list of relevant and
marginal chiral primary operators.  We have seen that $\Tr F^2$ and
$\Tr W_\alpha W^\alpha$ are always dimension 4 and dimension 3 operators
in these theories since they transform in the ${\bf 1}$ of
$SU(3)$. The trilinear terms that we found in the
decomposition of the ${\bf 50}$ in eq. (\ref{yuk1}) are of particular interest.
We saw  that since the ${\bf 10}$ of $SU(3)$ is
invariant under $Z_3$ there were  Kaluza-Klein states that were not projected
out and coupled to chiral primary operators on the boundary.
Under the
group action of $\Delta(3)$ one component of the
${\bf 10}$ will be invariant\footnote{This invariance
is familiar from
the $SU(3)_{\rm flavor}$ symmetry of the quark model: the baryon octet has
two states, $\Lambda$ and $\Sigma^0$, that are invariant under cyclic
permutations of the  three flavors, while the baryon decuplet has one
such state, $\Sigma^0$.}.
A brief inspection of Table I of  \cite{discrete}
shows that the all the subgroups $\Delta(3n^2)$
will preserve these Kaluza-Klein states, while the other non-Abelian
groups ($\Delta(6n^2)$ and $\Sigma(m)$) will not.

\section{${\cal N}=2$ Supersymmetric Theories}

In \cite{KS,LNV} ${\cal N}=2$ SCFTs were constructed by studying D3 branes at
orbifold singularities
of the form $R^4/\Gamma$, where $\Gamma \subset SU(2)$ is a discrete group.
The groups $\Gamma$ fall into the ADE classification.
As in the  ${\cal N}=1$ case,
the worldvolume theory is constructed by taking $N|\Gamma|$ D3 branes on the
covering space
and performing projection $\Gamma$ on the worldvolume fields and the Chan Paton
factors. Again
conformal field theories are expected when  the representation of $\Gamma$
acting
on the Chan Paton factors
is chosen to be the $N$-fold copy of the regular representation, which
 translates to the study of Type IIB string
theory on
an orbifold of $AdS_5 \times S^5$ where the orbifold group acts only on the
$S^5$ factor.
The $SO(4,2)$ isometry of $AdS_5$  is not broken and  corresponds to the
conformal symmetry of the SCFT on the D3 branes worldvolume.
The $SU(4)$ isometry of the sphere is broken to  $SU(2)_R \times U(1)_R\times
\Gamma$.
The $SU(2)_R \times U(1)_R$ factor  is the $R$-symmetry
 of the boundary ${\cal N}=2$ D3
brane theory.
The $\Gamma$ factor is
 a  discrete global symmetry.

This orbifold acts only on four of the six coordinates transverse to the D3
branes worldvolume
and there is a fixed plane of its action.
This implies that the supergravity description is not valid even at large $N$.
Nevertheless we will perform the analysis of the Kaluza-Klein spectrum and
relate it to
chiral primary operators of the boundary theory.
The analysis suggests that chiral information is still encoded in the
supergravity description.

Consider first the $A_{n-1}$ case. The discrete group $\Gamma = Z_n$ acts
as
\beqar
X^1 &\rightarrow& e^{\frac{2 \pi i}{n}} X^1 \non\\
X^2 &\rightarrow& e^{\frac{-2 \pi i}{n}} X^2
\label{zn}
\stop
\eeqar

The gauge group, global symmetries and field content of the D3 brane
theory is given in Table 2.
\begin{table}[htbp]
\centering
\begin{tabular}{cccccc|c}
 & $SU(N)_1$ & $SU(N)_2$  & $SU(N)_3$ & ... &  $SU(N)_n$  & $U(1)_R$
\\
\hline
$Q_1$ & \Yfund & $\overline{\Yfund}$ & {\bf 1}  & ... &  {\bf 1} & $0 \vbr$ \\
$\tilde Q_1$ & $\overline{\Yfund}$ & \Yfund & {\bf 1}  & ... &  {\bf 1} & $0
\vbr$ \\
$Q_2$ & {\bf 1} & \Yfund &  $\overline{\Yfund}$ & ...
& {\bf 1}  & $0 \vbr$ \\
$\tilde Q_2$ & {\bf 1} & $\overline{\Yfund}$ &  \Yfund & ...
& {\bf 1}  & $0 \vbr$ \\
\vdots &&&&&& \\
$Q_n$ & $\overline{\Yfund}$ & {\bf 1} & {\bf 1} & ...
&  $\Yfund$  &  $0 \vbr$ \\
$\tilde Q_n$ & \Yfund & {\bf 1} & {\bf 1} & ...
&  $\overline{\Yfund}$  &  $0 \vbr$ \\
$\Phi_1$ &Ad & {\bf 1} & {\bf 1}  & ... & {\bf 1}
 & $2 \vbr$ \\
\vdots &&&&&& \\
$\Phi_n$ & {\bf 1}  & {\bf 1} & {\bf 1}  & ... &Ad
& $2 \vbr$ \\
\end{tabular}
\label{N=2}
\parbox{4in}{\caption{Field content of the ${\cal N}=2$ theory, the $SU(2)_R$
symmetry is not manifest in the ${\cal N}=1$ notation used here.}}
\end{table}

First we consider  the particles in the family  (\ref{k1}).
We explicitly check the relevant and marginal operators for $k=2,3,4$ below.
For $k=2$, decomposing the ${\bf 20^\prime}$ into representations of
$SU(2) \times SU(2)_R \times U(1)_R$ gives:
\beq
{\bf 20^\prime} = ({\bf 1},{\bf 1})_0 +  ({\bf 1},{\bf 1})_4 +  ({\bf 1},{\bf
1})_{-4} +
 ({\bf 2},{\bf 2})_2  +  ({\bf 2},{\bf 2})_{-2} + ({\bf 3},{\bf 3})_0 \stop
\eeq
If we now perform the $Z_n$ projection\footnote{The $Z_n$ acts on the
${\bf 2}$ of $SU(2)$ as
$(x^1,x^2) \rightarrow (e^{\frac{2 \pi i}{n}} x^1, e^{-\frac{2 \pi i}{n}}
x^2)$.}
 of the $SU(2)$ we find that the
following components are invariant (labeled by $SU(2)_R \times
U(1)_R$):
\beq
{\bf 1}_0 + {\bf 1}_4 + {\bf 1}_{-4} + {\bf 3}_0 \stop
\label{components}
\eeq
At this point we need to identify the superconformal $R$-charge.  To do this
we make use of the fact that in ${\cal N}=1$ language there is an additional
$R$-symmetry, $U(1)_J$, which is a subgroup of  $SU(2)_R$.  Under this
symmetry adjoint chiral multiplets, $\Phi$, have charge 0, while fundamental
chiral multiplets, $\tilde Q$ and $Q$,  have charge 1.
We can then identify the superconformal $R$-charge as:
\beq
R_{\rm sc} = {{1}\over{3}} R + {{2}\over{3}} J \stop
\eeq
One can check that this gives the correct charge assignments to the gauginos
and the scalars (charges 1 and 2/3 respectively).
The corresponding $R_{\rm sc}$-charges of the  components in
(\ref{components}) are (0,4/3,-4/3,4/3).
Thus ${\bf 1}_4$ and ${\bf 3}_0$ have the
the correct $R_{\rm sc}$-charges to couple to dimension 2
chiral primary operators.  The ${\bf 1}_4$  Kaluza-Klein mode can be be
identified by its
quantum numbers as the coupling to the $Z_n$ invariant chiral primary operator
$\sum_{i=1}^n \Tr \Phi^2_i$.
The  ${\bf 3}_0$ Kaluza-Klein mode has to couple to a $Z_n$ invariant chiral
primary operator
with charges like
$\tilde Q Q$.
The chiral primary operator is identified\footnote{We would like to thank M. Schmaltz for pointing this out to us.}
 as 
 $\sum_{i=1}^n\tilde Q_{i}Q_{i}$.
Note this chiral primary
 operator vanishes due to the
$F$-term equations for
$U(N)$ gauge groups. 
We can also see that since the ${\bf 1}_4$ came from the
$ ({\bf 1},{\bf 1})_4$, it will be invariant under any projection $\Gamma$ that
preserves ${\cal N}=2$ supersymmetry, so $\sum_i \Tr \Phi_i^2$ will be a
dimension 2
chiral primary operator in any such theory.

For $k=3$ decomposing the ${\bf 50}$ into representations of $SU(2) \times
SU(2)_R \times
U(1)_R$ gives:
\begin{eqnarray}
{\bf 50} &=& ({\bf 1},{\bf 1})_2 +  ({\bf 1},{\bf 1})_{-2} +
({\bf 1},{\bf 1})_{6} +  ({\bf 1},{\bf 1})_{-6}+
 ({\bf 2},{\bf 2})_0  + ({\bf 2},{\bf 2})_4  + ({\bf 2},{\bf 2})_{-4}
\nonumber \\
&& + ({\bf 3},{\bf 3})_2 + ({\bf 3},{\bf 3})_{-2} + ({\bf 4},{\bf 4})_0  \stop
\end{eqnarray}
The $Z_n$ projection leaves invariant
\beq
 {\bf 1}_2 +  {\bf 1}_{-2} +  {\bf 1}_{6}  +  {\bf 1}_{-6}+ {\bf 3}_2 + {\bf
3}_{-2}   \stop
\eeq
The corresponding $R_{\rm sc}$-charges  are (2/3,-2/3,2,-2,2,2/3), thus
 ${\bf 1}_{6}$ and  ${\bf 3}_2$ have the
correct $R_{\rm sc}$-charge to couple to a dimension 3
chiral primary operator.  From their
quantum numbers we see that they couple to $\sum_{i=1}^n\Tr \Phi_i^3$ and
 $\sum_{i=1}^n\left(\tilde Q_{i-1} \Phi_i Q_{i-1} - \tilde Q_{i} \Phi_i Q_{i}
\right)$.
The choice of the latter is dictated by the need to remove derivatives of the
superpotential in order to get a
primary operator.
Since the  ${\bf 1}_{6}$ came from $({\bf 1},{\bf 1})_{6}$ we see  that
$\sum_{i=1}^n\Tr \Phi_i^3$
will
be
a dimension 3 operator in any theory obtained by a projection that preserves
${\cal N}=2$.

For $k=4$ decomposing the ${\bf 105}$ into representations of
$SU(2) \times SU(2)_R \times U(1)_R$ gives:
\begin{eqnarray}
{\bf 105} &=&  ({\bf 1},{\bf 1})_0 +  ({\bf 1},{\bf 1})_4 +  ({\bf 1},{\bf
1})_{-4} +  ({\bf 1},{\bf 1})_{8}
+  ({\bf 1},{\bf 1})_{-8}+
 ({\bf 2},{\bf 2})_2  +  ({\bf 2},{\bf 2})_{-2}   \nonumber \\
&& + ({\bf 2},{\bf 2})_6  + ({\bf 2},{\bf 2})_{-6}   +   ({\bf 3},{\bf 3})_0 +
({\bf 3},{\bf 3})_4
+ ({\bf 3},{\bf 3})_{-4} + ({\bf 4},{\bf 4})_2  + ({\bf 4},{\bf 4})_{-2}
\nonumber \\
&& + ({\bf 5},{\bf 5})_0 \stop
\end{eqnarray}
The $Z_n$ projection leaves invariant
\beq
  {\bf 1}_0 +{\bf 1}_4 +  {\bf 1}_{-4} +  {\bf 1}_{8}  +  {\bf 1}_{-8}  + {\bf
3}_0 + {\bf 3}_4 + {\bf 3}_{-4}   + {\bf 5}_{0}   \stop
\eeq
The corresponding $R_{\rm sc}$-charges  are
(0,4/3,-4/3,8/3,-8/3,4/3,8/3,0,8/3), thus
 ${\bf 1}_{8}$,  ${\bf 3}_4$, and  ${\bf 5}_0$ have the
correct $R_{\rm sc}$-charge to couple to a dimension 4
chiral primary operator.  From their
quantum numbers we see that they couple to  $\sum_{i=1}^n\Tr \Phi_i^4$,
 $\sum_{i=1}^n\left(\tilde Q_{i-1} \Phi_i^2 Q_{i-1} - \tilde Q_{i} \Phi_i^2
Q_{i} \right)$ and
 $\sum_{i=1}^n\left(\tilde Q_{i-1}Q_{i-1} - \tilde Q_{i}Q_{i} \right)^2$ respectively\footnote{Note 
that one can 
recast the chiral primary operators in a different form by adding appropriate powers of lower dimension
chiral primary operators.}. We see again that $\Tr \Phi^4$ will be
a dimension 4 operator for any choice of projection that preserves
${\cal N}=2$.
In general we expect Kaluza-Klein modes with masses (\ref{k1}) will
 couple  to dimension $k$ chiral primary operators in the boundary SCFT.

Next we consider particles in the family  (\ref{k2}).
To explicitly check the relevant chiral
operator for $k=0$ we decompose the ${\bf 10}$
into representations of $SU(2) \times SU(2)_R \times U(1)_R$:
\beq
{\bf 10} =  ({\bf 2},{\bf 2})_0  +  ({\bf 3},{\bf 1})_{-2} +
({\bf 1},{\bf 3})_{2} \stop
\eeq
The  $Z_n$
projection leaves invariant
\beq
 {\bf 1}_{-2} +{\bf 3}_{2} \stop
\eeq
The corresponding $R_{\rm sc}$-charges  are (-2/3,2), thus
$ {\bf 3}_{2} $ has the
correct $R_{\rm sc}$-charge to couple to
a dimension 3 chiral primary operator.  The corresponding operator is
$\sum_{i=1}^n\Tr W_{\alpha}^iW_i^{\alpha}$, and we see that it will be
dimension

3 with any projection that preserves ${\cal N}=2$.

For $k=1$ we decompose the ${\bf  45}$ as:
\begin{eqnarray}
{\bf 45} &=&  ({\bf 2},{\bf 2})_2  +  ({\bf 2},{\bf 2})_{-2}  +
({\bf 3},{\bf 1})_{-4} + ({\bf 1},{\bf 3})_{4}
+({\bf 3},{\bf 1})_{0} + ({\bf 1},{\bf 3})_{0} + ({\bf 3},{\bf 3})_{0}
\nonumber \\
&&+  ({\bf 4},{\bf 2})_{-2}  +  ({\bf 2},{\bf 4})_{2} \stop
\end{eqnarray}
The  $Z_n$
projection leaves invariant
\beq
{\bf 1}_{-4} +{\bf 3}_{4} +{\bf 1}_{0} +{\bf 3}_{0} +{\bf 3}_{0} \stop
\eeq
The corresponding $R_{\rm sc}$-charges  are (-4/3,8/3,0,4/3,4/3), thus
the ${\bf 3}_{4}$ has the
correct $R_{\rm sc}$-charge to couple to
a dimension 4 chiral primary operator. The corresponding operator is
$\sum_{i=1}^n\Tr W_{\alpha}^iW^{\alpha}_i \Phi_i$.  Again this result is
completely
general as long as ${\cal N}=2$ supersymmetry is preserved.
In general
we expect Kaluza-Klein modes with masses (\ref{k2})
to couple to dimension $k+3$ chiral primary operator in the boundary SCFT.

Now consider  particles in the family (\ref{k3}).
For $k=0$ we get the relevant operator $\sum_{i=1}^n\Tr F_i^2$.  Again the
$Z_n$
projection
on
the only relevant representation, ${\bf 1}$, is trivial, a result that holds
independent of the projection $\Gamma$.
We expect Kaluza-Klein modes with
masses (\ref{k3})
to couple to dimension $k+3$ chiral primary operator in the boundary SCFT.

Finally consider the spin one family  (\ref{vec}).
The only relevant mode is the massless
mode which is in the ${\bf 15}$ of $SU(4)$.  Decomposing the ${\bf 15}$
we find
\beq
{\bf 15} =  ({\bf 1},{\bf 1})_0 +  ({\bf 2},{\bf 2})_2  +  ({\bf 2},{\bf
2})_{-2} +  ({\bf 3},{\bf 1})_{0}
+ ({\bf 1},{\bf 3})_{0} \stop
\eeq
The  $Z_n$
projection leaves invariant
\beq
{\bf 1}_{0} +{\bf 1}_{0} +{\bf 3}_{0}  \comma
\eeq
corresponding to the currents of the unbroken $SU(2)_R \times U(1)_R$
symmetry and
one extra $Z_n$ invariant current which is a remnant of the broken
$SU(2)$ symmetry.  As above the the currents corresponding to the
unbroken $R$-symmetry will be dimension 3 under any projection that
preserves ${\cal N}=2$ supersymmetry.

It is straightforward to extend the analysis to the $D_n$ and $E_n$ orbifolds.
Since $Z_n$ is
a subgroup of  $D_n,E_6$, and $E_7$, we further restrict the list of relevant
and
marginal chiral primary operators. Also, the form of the generators of the
$E_8$ singularity
 (Icosahedral group) \cite{slov}
dictates
the same list of
operators as $E_6$ and $E_7$.  In going from the $A_{n-1}$  to $D_n$ we
find that
the ${\bf 3}$ no longer has an invariant component, while the ${\bf 5}$ still
does.
So there are no analogs of  the chiral primary operators:
$\tilde Q_i \Phi_i Q_i - \tilde Q_{i-1} \Phi_i Q_{i-1}$, and
$\tilde Q_i \Phi_i^2 Q_i - \tilde Q_{i-1} \Phi_i^2 Q_{i-1}$. Under the $E_n$
projections
neither the ${\bf 3}$ nor the ${\bf 5}$ have invariant components, so we also
do not have an analog of the
$\left(\tilde Q_{i-1}Q_{i-1} - \tilde Q_{i}Q_{i} \right)^2$ operator.

\section{Nonsupersymmetric Theories}

Examples of  ${\cal N}=0$ candidates for CFTs in the large $N$ limit
can be constructed
by considering
orbifold singularities
of the form $R^6/\Gamma$, where $\Gamma \subset SU(4)$ is a discrete subgroup,
and
 the orbifold group acts only on the
$S^5$ factor of  $AdS_5 \times S^5$.
As before,
the $SO(4,2)$ isometry of $AdS_5$  is not broken and  corresponds to the
conformal symmetry of the boundary CFT, while the
the $SU(4)$ isometry of $S^5$  is broken to $\Gamma$
which becomes a discrete global symmetry of the CFT.

Consider for example  the $Z_5$ orbifold example of \cite{KS}
\beqar
X^1 &\rightarrow& e^{\frac{2 \pi i}{5}} X^1 \non\\
X^2 &\rightarrow& e^{\frac{6 \pi i}{5}} X^2
\label{zn0}
\stop
\eeqar
The gauge group, global symmetries and field content of the D3 brane
theory is given in Table 3.

\begin{table}[htbp]
\centering
\begin{tabular}{cccccc}
 & $SU(N)_i$ & $SU(N)_{i+1}$  & $SU(N)_{i+2}$ & $SU(N)_{i-2}$  & $SU(N)_{i-1}$

\\
\hline
$\phi_{i,i+1}$ & \Yfund & $\overline{\Yfund}$ & {\bf 1}  & {\bf 1} & {\bf 1}
$\vbr$
\\
$\phi_{i,i+2}$  & \Yfund & {\bf 1} &  $\overline{\Yfund}$ & {\bf 1} & {\bf 1}
$\vbr$ \\
$\phi_i$ & Ad & {\bf 1} & {\bf 1}  & {\bf 1} & {\bf 1}   $\vbr$ \\
$\psi_{i,i+1}$ & \Yfund & $\overline{\Yfund}$ & {\bf 1}  & {\bf 1} & {\bf 1}
$\vbr$ \\
$\psi_{i,i-1}$ & \Yfund  & {\bf 1}  & {\bf 1} & {\bf 1}  & $\overline{\Yfund}$
$\vbr$ \\
$\psi_{i,i+2}$  & \Yfund & {\bf 1} &  $\overline{\Yfund}$ & {\bf 1} & {\bf 1}
$\vbr$ \\
$\psi_{i,i-2}$  & \Yfund & {\bf 1}  & {\bf 1} &  $\overline{\Yfund}$ & {\bf 1}
$\vbr$ \\
\end{tabular}
\label{N=0}
\parbox{4in}{\caption{Field content of the ${\cal N}=0$ theory, where $Z_5$
cyclicly
permutes the gauge groups. $\phi$'s are scalars and $\psi$'s are fermions.
We have not listed the conjugates of the
bifundamental fields.}}
\end{table}

As in the $\cN=2$ case the orbifold group has a fixed plane and we do not
expect the supergravity description
to be valid even at large $N$.
Unlike the supersymmetric cases we also do not a priori expect the masses of
the Kaluza-Klein harmonics to be
protected from quantum and string corrections.
However, there is still the possibility that dimensions of certain primary
operators in the the $\cN=0$ boundary CFT do not receive quantum corrections,
and that the
$AdS/CFT$ correspondence generalized to the $\cN=0$ theories will imply that
masses of
certain Kaluza-Klein modes do not receive quantum corrections.
With this in mind we will now carry out some examples of an analysis similar
to that of the previous sections.
In order to get the Kaluza-Klein harmonics on the supergravity side we will
project those of $AdS_5 \times S^5$
on $Z_5$ invariant states\footnote{The $Z_5$ acts on the ${\bf 4}$ of $SU(4)$
as
$(x^1,x^2,x^3,x^4) \rightarrow
(e^{\frac{2 \pi i}{5}} x^1, e^{\frac{-2 \pi i}{5}}
x^2, e^{\frac{4 \pi i}{5}} x^3, e^{\frac{-4 \pi i}{5}}  x^4)$.}.
The results are summarized in Table 4.

In order to construct the primary operators of the $\cN=0$ boundary CFT we have
to remove dependences
on the derivatives with respect to the fields of the Yukawa and  quartic
couplings.
We expect Yukawa couplings for each triangle with two fermion lines and one
scalar
line in the quiver diagram description of the model, and
quartic couplings for each square with four scalar lines \cite{KS,LNV}.

Consider the invariant Kaluza-Klein states.
There are four invariant states from the $k=2$ modes in (\ref{k1}) that
transform in the  ${\bf 20'}$
of $SU(4)$ and should couple to dimension 2 primary operators.
Since the ${\bf 20'}$ is made of two ${\bf 6}$'s of $SU(4)$ we should construct
primary operators bilinear in the
scalars. We have $\Tr \phi_i^2, \Tr \phi_{i,i+1}\phi_{i+1,i}, \Tr
\phi_{i,i+2}\phi_{i+2,i}$ where
$\phi_i$ is the adjoint scalar of the $i$-th gauge group and $ \phi_{i,j}$ is
the scalar associated with the line
connecting the $i,j$ nodes of the quiver diagram.
The above operators (and also in the other examples below)
have an implicit summation on the different nodes $i$ in order to make them
$Z_5$
invariant.
We do not seem to see a possible fourth primary operator of dimension 2 in
the theory. This suggests that the number of invariant Kaluza-Klein states is
larger than the number
of primary operators in the boundary CFT. This is not too surprising in view of
the fact that
we lack here the chirality constraint which was important in
the supersymmetric case. We will see more examples of this in the following.

There are 10 invariant states from the $k=3$ modes in (\ref{k1}) that
transform in the  ${\bf 50}$
of $SU(4)$ and should couple to dimension 3  primary operators.
Since the ${\bf 50}$ is made of three ${\bf 6}$'s of $SU(4)$ we should
construct
primary operators from three
scalars. An obvious one is $\Tr \phi_i^3$.
Others can are made of  $\phi_{i,i+1},\phi_{i+1,i},\phi_{i,i+2},\phi_{i+2,i}$.
Also in this case we have more
invariant Kaluza-Klein states  than the number
of primary operators in the boundary CFT.
Similarly there are 21 invariant states from the
 $k=4$ modes in (\ref{k1}) that transform in the  ${\bf 105}$
of $SU(4)$ and should couple to dimension 4 primary operators.
Since the ${\bf 105}$ is made of four  ${\bf 6}$'s of $SU(4)$ we should
construct
primary operators from four
scalars. One of them is  $\Tr \phi_i^4$ and as before the rest are made from
the other scalars.
Clearly we have more invariant Kaluza-Klein modes than primary operators.

Consider now the invariant Kaluza-Klein states of (\ref{k2}).
There are two invariant states from the $k=0$ modes that transform in the
${\bf 10}$
of $SU(4)$ and should couple to dimension 3 primary operators.
Since the ${\bf 10}$ is made of two ${\bf 4}$'s of $SU(4)$ we should construct
primary operators bilinear in the
fermions. There are two
such independent primary  operators $\Tr \psi_{i,i+1}\psi_{i+1,i}, \Tr
\psi_{i,i+2}\psi_{i+2,i}$
 where
 $\psi_{i,j}$ is the fermion  associated with the line
connecting the $i,j$ nodes of the quiver diagram.
There are 9 invariant states from the $k=1$ modes that transform in the
${\bf 45}$
of $SU(4)$ and should couple to dimension 4  primary operators.
Since the ${\bf 45}$ is made of two ${\bf 4}$'s and a  ${\bf 6}$ of $SU(4)$ the
 primary
operators are bilinear in the
fermions and linear in the scalars and again we seem to see that there are
 more invariant Kaluza-Klein modes than primary operators.

The  dilaton $k=0$ in (\ref{k3}) and the graviton $k=0$ in (\ref{grav})
 are not projected out and
couple to the relevant operators $\Tr F^2$ and $T_{\mu\nu}$.

\section{Summary and Discussion}

\begin{table}[htbp]
\centering
\begin{tabular}{ccc|ccccccc}
 &  &  $\Gamma$ & $A_n$  & $D_n$& $E_n$& $Z_3$ & $\Delta(3n^2)$ & $ \ne
\Delta(3n^2)$ &$Z_5$ \\
 &  &${\cal N}$ & 2 & 2& 2& 1  & 1 & 1 & 0 \\
 ${\cal R}$ &spin  &$\Delta$  &&&& & & &  \\
\hline
${\bf 20^\prime}$ & 0 & 2  & ${\bf 1}_{4/3}$,${\bf 3}_{4/3}$ & ${\bf 1}_{4/3}$
& ${\bf 1}_{4/3}$& - & -  & - & 4   \\
{\bf 50}  & 0 & 3 & ${\bf 1}_2$,${\bf 3}_2$  & ${\bf 1}_2$& ${\bf 1}_2$&
$10_2$
&  $1_2$  & - & 10 \\
{\bf 105} & 0 & 4 &  ${\bf 1}_{8/3}$,${\bf 3}_{8/3}$,${\bf 5}_{8/3}$   &${\bf
1}_{8/3}$,${\bf 5}_{8/3}$ &${\bf 1}_{8/3}$& - &  -& - & 21 \\ \hline
{\bf 10} & 0 & 3 & ${\bf 3}_2$   &${\bf 3}_2$&${\bf 3}_2$& $1_{2}$
& $1_{2}$  & $1_{2}$ & 2 \\
{\bf 45} & 0 & 4  & ${\bf 3}_{8/3}$ &${\bf 3}_{8/3}$&${\bf 3}_{8/3}$& -
& -   & -  & 9 \\ \hline
{\bf 1}  & 0 & 4 & ${\bf 1}_0$  &${\bf 1}_0$&${\bf 1}_0$&   $1_0$
&   $1_0$ &   $1_0$ & $1$ \\ \hline
{\bf 15}  & 1 & 3 & ${\bf 1}_0$,${\bf 1}_0$,${\bf 3}_0$  & ${\bf 1}_0$,${\bf
3}_0$ & ${\bf 1}_0$,${\bf 3}_0$&$1_0$, $8_0$
&$1_0$, $1_0$, $1_0$  &$1_0$ &  3 \\
\end{tabular}
\label{summary}
\parbox{4in}{\caption{Kaluza-Klein harmonic projections. The projections are
labeled
by $\Gamma$ and ${\cal N}$, while for the $SU(4)$ representations, the spin and
scaling dimensions are
also listed.  For ${\cal N}=2$ the invariant components of the representations
are
labeled by $SU(2)_R \times U(1)_{R_{\rm sc}}$, for ${\cal N}=1$ by the number
of
components and $U(1)_{R_{\rm sc}}$, and for ${\cal N}=0$ by the number of
components.  The notation $ \ne
\Delta(3n^2)$ refers to the other non-Abelian subgroups of $SU(3)$,
$\Delta(6n^2)$ and $\Sigma(m)$.}}
\end{table}

In this work we studied the relation between (chiral) primary operators of
(super) conformal field theories
in four dimensions constructed in \cite{KS,LNV}
 and the Kaluza-Klein states of supergravity on orbifolds of $AdS_5 \times
S^5$.
This generalizes the relation between the chiral primary operators of $\cN=4$
SCFT and
 Kaluza-Klein states of supergravity on $AdS_5 \times S^5$ found in
\cite{witten-one}.
We obtained the Kaluza-Klein modes in the orbifold models by projecting those
of
 supergravity on $AdS_5 \times S^5$
on $\Gamma$ invariant states where $\Gamma$ is the orbifold group.
In Table 4 we summarize the results.
In the $\cN=0$ we saw more Kaluza-Klein states than primary
operators.

Note that even in cases where the supergravity description is not applicable
we see that chiral information is still reliably encoded in this description.
The fact that BPS information is obtained correctly even when the
supergravity description is not valid is already a known phenomenon.
For instance, when considering gauge theories via wrapping the fivebrane 
of eleven dimensional  supergravity (M theory)  
on a Riemann surface in order to obtain $N=2$ 
supersymmetric  gauge theories in four dimensions \cite{wbrane}
there are points in the $N=2$
moduli space where the Riemann surface  degenerates and the eleven 
dimensional 
supergravity description is not valid. Nevertheless the spectrum of BPS
particles is obtained correctly.
The basic reason is that the BPS spectrum is protected from quantum 
corrections and the BPS mass formula continues to hold even when 
extrapolated to regions where supergravity theory does not
provide  a good description.
Similar phenomenon occurs
 for $N=1$ supersymmetric  gauge theories in four dimensions
that are 
obtained via wrapping the fivebrane on a Riemann surface \cite{hoo,wqcd}. 
In our case we see another example of this phenomenon, since the spectrum of
chiral operators is protected from quantum corrections.
A deeper analysis of this is still lacking.

The results of this work  can be generalized in a straightforward way to
orbifolds of  $AdS_7 \times
S^4$
and $AdS_4 \times S^7$ that lead to six and three dimensional SCFTs
respectively \cite{FKPZ,Gomi}.

Note that in our analysis we have not seen Kaluza-Klein modes that correspond
in the boundary SCFTs to the Yukawa couplings that arise from the
superpotential of the $\cN=4$ theory upon projection.
The reason being that these Yukawa couplings do not correspond to  primary
operators.
They should however play an important role.
The orbifold theories  have a vanishing one-loop $\beta$-function
\cite{KS,LNV}. If the  Yukawa couplings vanish then
the two-loop $\beta$-function will not not be zero, in fact the two-loop
$\beta$-function
will be positive and these theories will be infrared free theories, that is,
they will be conformal but trivial.
Thus, it would be important to carefully analyze these couplings and their
effect on the higher loop $\beta$-functions.

It is interesting to note that for the ${\cal N}=0,1$ cases we can easily see that
the fixed
line is only present in the large $N$ limit unless we modify the
Yukawa couplings with $N$  dependent corrections that vanish as $N\rightarrow \infty$\footnote{We 
would like to thank C. Vafa for a discussion on this point.
See also a discussion in \cite{BKV}.}. 
In the $N=1$ cases where we have a Leigh-Strassler type argument \cite{LS,KS} 
such a modification is guaranteed to exist.
Consider the $\cN=0$ case.
The vanishing of the one-loop
Yukawa $\beta$-function \cite{Machacek} works for a $U(N)$  gauge group because
there is
a cancelation between Yukawa contributions, which receive a counting factor $N$
from the $N$ fundamentals, and a gauge contribution which is proportional to
the
Casimir of the the fundamental, $N/2$.  Of course the $U(1)$ sub-group is
infrared-free,
so the fixed point can only occur for the $SU(N)$ theory, but then the
cancellation fails
at next-to-leading order in $1/N$ since the Casimir of $SU(N)$ is $(N^2-1)/2N$.
It would be important to show that a modification of the Yukawa couplings and a fixed line
at finite $N$ exist also in the
$\cN=1$ theories.

\newpage

\section*{Acknowledgements}

We would like to thank O. Aharony, K. Bardacki, J. de Boer, S.  Kachru,
A. Lawrence,  M. Schmaltz,  A. Uranga, C. Vafa and  Z. Yin, for discussions.
We would also like to thank S.  Kachru for comments on the manuscript.
 This work
was
supported in part by NSF
grant PHY-951497 and DOE grant DE-AC03-76SF00098.

\end{document}